\documentclass[preprint]{revtex4}
\usepackage{bm}
\input{epsf}
\begin{document}
\title{Electron and Phonon Confinement and Surface 
Phonon Modes in CdSe-CdS Core-Shell Nanocrystals}

\author{A. Singha$^{\dagger}$, B. Satpati$^{\ddagger}$, P.V. Satyam$^{\ddagger}
$ and Anushree Roy$^{\dagger}$ }
\email{anushree@phy.iitkgp.ernet.in}
\affiliation{$^{\dagger}$Department of Physics,
Indian Institute of Technology, Kharagpur 721 302, WB, India\\
$^{\ddagger}$Institute of Physics, Bhubaneswar  751005, India}

\begin{abstract} 
Optical and vibrational properties of bare and CdS 
shelled CdSe nanocrystalline particles are investigated. To confirm the 
formation of such nanocrystals in our 
samples we estimate their average particle sizes and size distributions 
using TEM measurements. From the line profile analysis of the images the 
core-shell structure in the particles has been confirmed. The blue shift 
in optical absorption spectra, analyzed using theoretical 
estimates based on the effective bond order model, establishes the 
electron confinement in the nanoparticles. Unique characteristics of the 
nanocrystals (which are absent in the corresponding bulk material), such 
as confinement of optical phonons and the appearance of surface phonons, 
are then discussed. Making use of the dielectric response function model 
we are able to match the experimental and theoretical values of the 
frequencies of the surface phonons. We believe that our studies using 
optical probes provide further evidence on the existence of core-shell 
structures in CdSe-CdS type materials.
\end{abstract}
\maketitle
\def\d{{\mathrm{d}}}

\section{Introduction} Over last couple of decades, unique physical
properties due to quantum confinement effects have been reported for a
wide range of semiconductor nanocrystalline materials \cite{Schmid:2003}.  
Since in these systems surface to volume ratio is high compared to that of
corresponding bulk materials, surface states play a crucial role in
determining their physical properties. The surface of nanocrystals is
made of atoms that are not fully coordinated.  Hence, they are highly
active and these surface atoms act like defect states unless they are
passivated by either organic ligands or higher bandgap semiconductor
materials \cite{Schmid:2003,Spanhel:1987}. Thus, the current direction in
this field of research includes modification of given size-quantized
semiconductor particles by means of surface chemistry. In the literature,
we find reports on quantum dot-quantum well (QDQW) and core-shell
nanocrystalline structures due to surface modifications of the particles.
QDQW is a three layered structure consisting of a size quantized particle
acting as a core [marked 'a' in Fig. 1(a)] and a complete layer of another 
material on the surface [marked 'b' in Fig. 1(a)] of this core, which is 
again covered by core material as an outermost shell. This 
core-shell-core structure is marked as a-b-a in Fig. 1(a). Colloquially, 
these particular structures are also
called 'nano-onions'.  Recently, Dorfs and Eychm\"{u}ller have outlined 
the preparation and characterization of a series of QDQW systems 
containing two wells \cite{Dorfs:2001}. Coating nanoparticles by another 
material yields core-shell nanocrystals [see Fig. 1(b)] \cite{Schmid:2003}.

\begin{figure}
\centerline{\epsfxsize=4.5in\epsffile{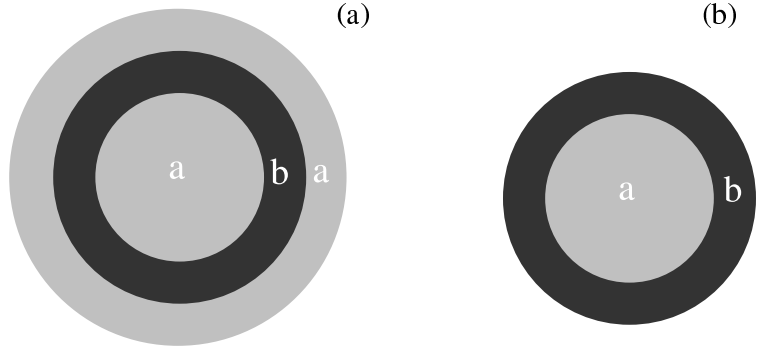}}
\caption{Schematic drawing of the nano-onion and core-shell type
nanoparticles}
\end{figure}

In core-shell strucure, like inorganic epitaxial shell growth, the organic
ligands cannot passivate both cationic and anionic surface sites of the
core \cite{Peng:1997}. The particles passivated by inorganic shell
structures are more robust than organically passivated nanocrystals and
have greater tolerance to processing conditions necessary for
incorporation into solid structures \cite{Wilson:1993}. For effective
surface passivation, the core particles having certain band gap are capped
with a higher band gap material.  Moreover, the conduction band energy of
the capping (shell) material is usually higher than that of the core
material with the valence band energy of the capping material being lower.
This energetic situation is called a type- I structure \cite{Schmid:2003}.
Due to presence of higher band gap capping material, the photogenerated
excitons in the core remain localized in the same region and are forced to
recombine while spatially confined in the core. Confinement of electrons
in the nanocrystals gives rise to blue shift in optical absorption and PL
spectra of the material. As the non-radiative decay channels through
surface states are not accessible for these electrons, core-shell
structures thus formed show higher luminescence quantum yields
\cite{Peng:1997,Spanhel:1987, Kortan:1990,Hoener:1992,
Mews:1994,Danek:1996,Hines:1996,Pradhan:2003} lower fluorescence life time
\cite{Hines:1996} and many other benefits \cite{Danek:1994} related to
tuning of band gap in two materials.

Like electrons, phonons are also confined in nanocrystals. In bulk
crystals, the phonon eigenstate is a plane wave and wavevector selection
rule for the first order Raman scattering requires $q \sim 0$. In
contrast, the spatial correlation function of the phonon becomes finite
due to its confinement in the nanocrystal and hence the $q \sim 0$ 
selection
rule gets relaxed. In general, the phonon-dispersion curves of bulk
crystals show the frequency $\omega$ to be a decreasing function of 
wavevector 
$q$. Hence, the first order Raman line shifts and broadens towards the low
frequency side for the nanocrystals. This has been
proposed and explained by phonon confinement model \cite{Campbell:1986}. 
Confinement of phonons
for Si, Ge, BN, CdS, CdS$_x$Se$_{1-x}$ and many other nanocrystals have
been reported in the literature
\cite{Richter:1981,Nemanich:1981,Tsu:1992,Sood:1992,Roy:1994,Roy:1996}.

As mentioned before, surface states play an important role in deciding
different physical properties of nanocrystals. For a plane wave
propagating in the x-direction in a bulk crystal, the temporal and spatial
variation of the wave is described by the factor $exp[i(kx-\omega t)]$,
where the wavevector $k=(\omega/c)\sqrt{\epsilon(\omega)}$;  
$\epsilon(\omega)$ is the dielectric constant of the crystal. In the
frequency range between bulk longitudinal optical (LO) phonon frequency,
$\omega_{LO}$, and transverse optical (TO) mode frequency, 
$\omega_{TO}$, $\epsilon(\omega) <0$, $k$ is imaginary. 
Therefore, in this frequency range
the wave decays exponentially in the medium, i.e. it can not propagate in 
bulk crystals and only surface modes exist \cite{Fuchs:1968}. Because of
enhanced surface to volume ratio these modes appear for nanocrystals and
it provides relevant information on surface states. Recently, 
Baranov et al have reported the surface 
phonon modes in ZnS shelled CdSe particles \cite{Baranov:2003}.

Characteristics of bare CdSe nanocrysals are now well established. CdSe
nanocrystals passivated with long organic chain (like TOPO) have room
temperature photoluminescence (PL) (quantum yield 10$\%$) with very long
fluorescence time \cite{Norris:1994,Hoheisel:1994}.  The inorganic
shelling of CdSe core by CdS has also been explored and shown to be better
for surface passivation. Previous HRTEM and XPS measurements have shown
that shell growth in this system does not form an alloy
\cite{Peng:1997,Nanda:1999}. The lattice mismatch of 3.9$\%$ between CdS 
and CdSe is small enough to allow epitaxial growth while still preventing
alloying. Moreover, the difference in band gap ( 2.42 eV for CdS and 1.74 
eV for CdSe at room temperature) is large enough for shell growth. The 
room temperature PL yield for CdSe-CdS core-shell system has been 
reported 
to be upto 50$\%$ \cite{Peng:1997}. The electronic structure in this 
material can be understood from molecular orbital model and particle in a 
box model, as discussed in  Ref. \cite{Peng:1997}. The same has also been 
discussed from the analysis of photoelectron spectra from core-shell 
structure \cite{Nanda:1999}.  The high
photostability of the system has been explained by confinement of
electrons and holes in the core-shell region. 

In this article, we have discussed the optical properties of
marcaptoacetic acid stabilized bare CdSe particles and CdS capped CdSe 
particles. The nanoparticles have been
characterized by their absorption spectra as well as by high resolution
transmission electron microscope (HRTEM) images.  The optical transition 
is modeled using
Effective bond order model (EBOM). Average 
particle size and size distribution of
the particles have been estimated from both HRTEM images and 
optical absorption measurements; which have been further supported by 
Raman measurements. 
We have quantitatively analyzed asymmetric Raman line shapes including
both confined optic modes and surface phonon modes. The observed
frequencies of the SP modes in these systems have been compared with the
calculations based on the dielectric response function model. In Section 
II of this article, we have briefly discussed the sample preparation 
procedure, which we have followed. Section III deals with analysis of 
HRTEM images of the samples. Section IV demonstrates the optical 
properties 
of the samples, studied by optical absorption and PL spectroscopy. Detail 
analysis of  phonon spectra is reported in Section V. Finally, in section 
VI, we have summarized our results with a few concluding remarks.

\section{Sample preparation} 

\subsection{Chemicals} 
For sample preparation, Cadmium 
perchlorate, [Cd(ClO$_4$)$_2$] and Marcaptoacetic 
acid [HS-CH$_2$-COOH] were purchased from Aldrich, USA and LOBA 
Chemie,India respectively. Anhydrous Sodium sulfite [Na$_2$SO$_3$] and 
Sodium sulfide [Na$_2$S] 
were from MERCK Ltd., India and  S. D. Fine Chem. LTD., India 
respectively. Solid Selenium powder[Se] and  Sodium hydroxide [NaOH]
were used as received.

\subsection{Preparation of stock solution of Na$_2$SeSO$_3$} 

1.0 gm of Se powder was added in 200 ml (1.1 M) hot Na$_2$SO$_3$ solution 
under stirring condition and then  boiled for  half an hour. Subsequently, 
the solution was cooled at room 
temperature and the aqueous layer was filtered using Whatman 
filter paper. This solution is used as the stock solution.
\centerline {Na$_2$SO$_3$ +Se $\rightarrow $ Na$_2$SeSO$_3$}

\subsection{Preparation of marcaptoacetic acid stabilized CdSe 
nanocrystals : Sample A}  

Following Liu {\em et al} \cite{Liu:2000}, 150 ml of Cd(ClO$_4$)$_2$ (2 x
10$^{-4}$ M) and 150 ml of HS-CH$_2$-COOH (2 x10$^{-4}$ M) were mixed and
stirred vigorously for about 5 min. Here, the latter acts as a stabilizing
agent. The pH of the solution was adjusted to 9.0 by adding aqueous NaOH
(1.0 M) solution and heated to 100$^o$C under N$_2$ atmosphere.
Subsequently, 0.5 ml of Na$_2$SeSO$_3$ was added drop wise to the solution
and boiled again for about half an hour under stirring condition. The 
solution turned orange in color indicating the formation of CdSe 
particles.

\subsection{Preparation of marcaptoacetic acid stabilized CdSe-CdS 
core-shell nanocrystals: Sample B} 

For the preparation of coated CdSe-CdS particle,
the above CdSe solution was cooled to 50$^o$C and 20 $\mu$l of 
Cd(ClO$_4$)$_2$
(0.1M) and 20 $\mu$l of Na$_2$S (0.1M) were added drop wise alternatively
under stirring condition in N$_2$ atmosphere. After 30 min of stirring the
solution turned orange-red in color due to formation of CdSe-CdS 
core-shell particles.

\section{Average particle size and size distribution : HRTEM}

\begin{figure}
\centerline{\epsfxsize=5.5in\epsffile{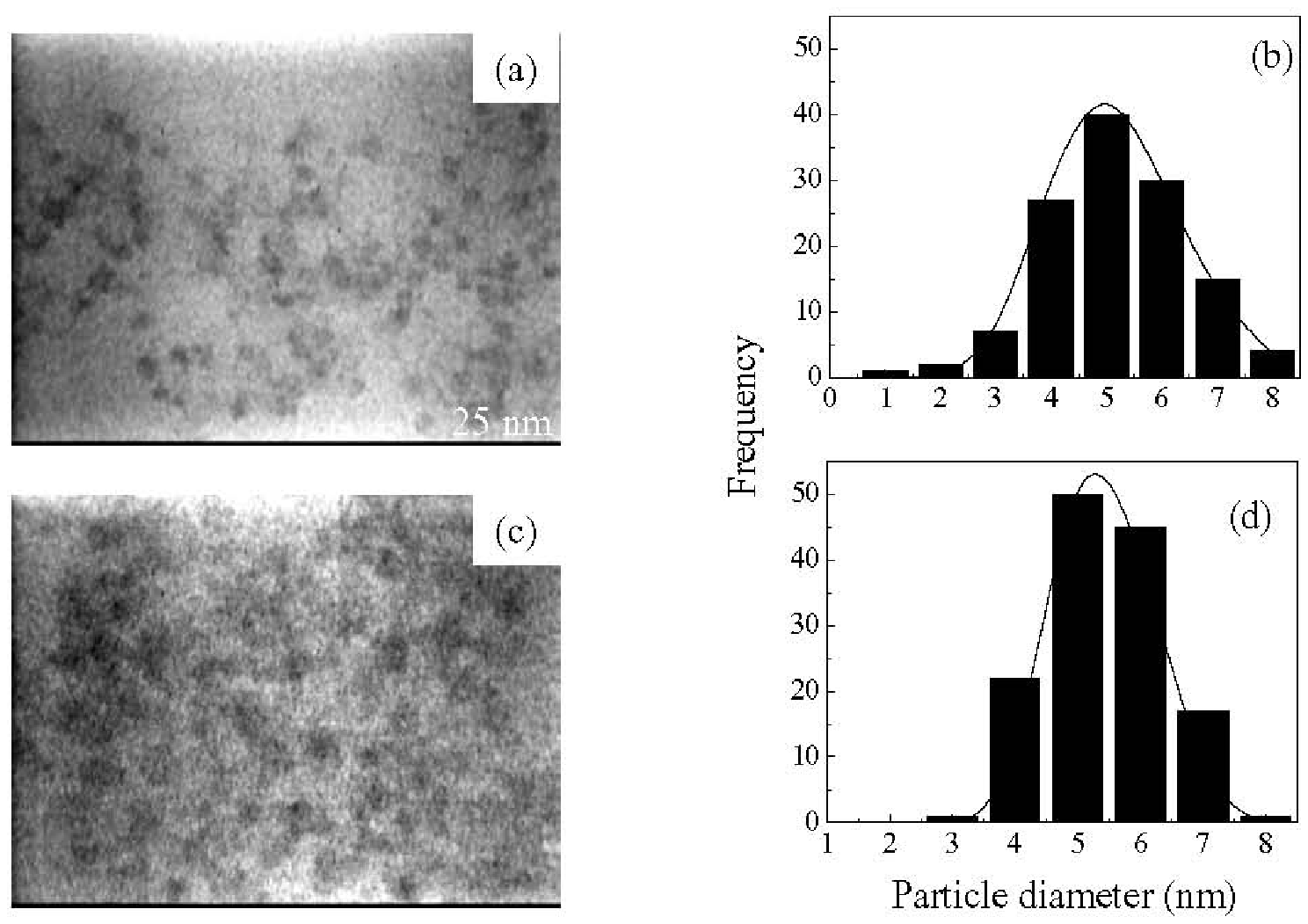}}
\caption{Low magnification micrographs and histogram of the nanocrystals. 
(a) is the TEM micrograph and (b) is the corresponding histogram for 
Samples A. Similarly, (c) is the micrograph and (d) is the corresponding 
histogram for Sample  B.}
\end{figure}   

Samples for Transmission Electron Microscopy  were deposited onto 300
mesh copper TEM grids coated with 50 nm carbon films.  Samples A and B are
suspended in water; thus, directly added drop-wise on the grid.
The excess
water was allowed to evaporate in air. The grids were examined in JEOL 
2010 microscope with Ultra-High Resolution (UHR)  microscope using a 
LaB$_6$ filament operated at 200 kV. HRTEM images of many
nanocrystals for each sample were measured and analyzed.

TEM has been used to determine the particle size and size distribution of the 
afore-mentioned nanocrsytals. Also, using the high
resolution TEM, it is possible to confirm the core-shell structure of 
nanocrystals \cite{Peng:1997}. Low magnification TEM micrographs are 
shown in Fig. 2 (a) and (c) for Sample A and B, respectively. Fig. 2(b) 
and (d) are the histogram plots,
obtained by measuring sizes of many particles per sample. Size
distribution for the nanoparticles are usually found to be log-normal:

\begin{equation}
P(d)=\frac{1}{\sqrt{2\pi}d\sigma}exp\left(-\frac{ln
(d/\bar{d})}{\sqrt{2}\sigma}\right)^{2}
\end{equation}

\noindent
Here $\bar{d}$ is the average size and $\sigma$ is related to the
size distribution of the particles. By fitting the frequency plot using
Eqn. 1 (solid lines in Fig. 2), we have estimated the average particle 
size ($\bar{d}$) and $\sigma$ of the particles, which are listed in Table 
I. The 
size distribution of the particles is estimated to be  more spread out for
bare particles (Sample A) compared to the  other sample.

\begin{figure}
\centerline{\epsfxsize=5.5in\epsffile{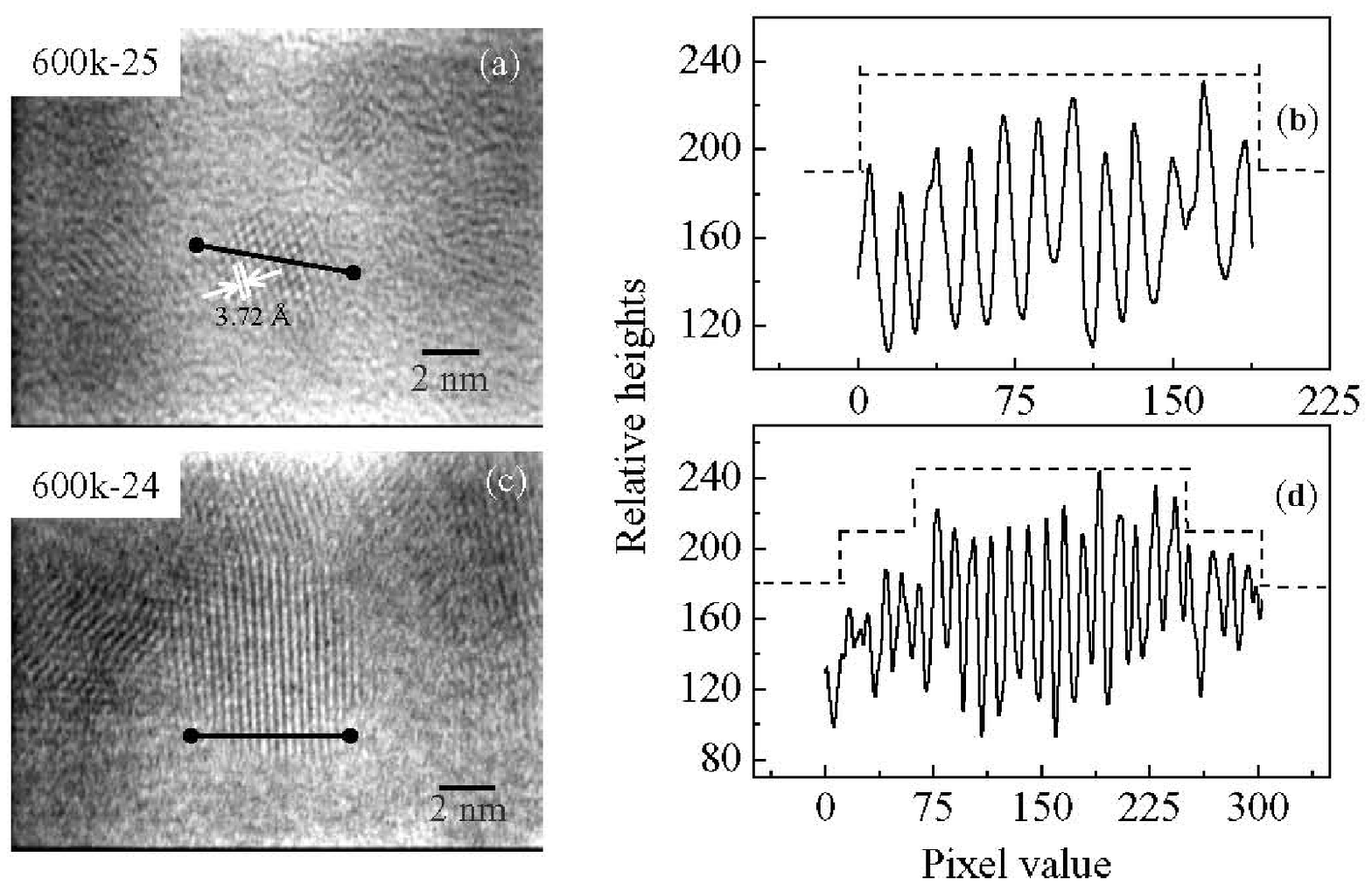}}
\caption{High resolution TEM micrographs and corresponding line profiles 
for Sample A and B.}
\end{figure} 

Transmission electron microscopy measurements provide the
convincing proof of the crystalline core-shell structure of the sample. 
Fig. 3 shows high resolution micrograph of selected nanocrystals along 
the line profiles. The contrast in the image depends on the
electron density in the object forming the image. Hence, CdS is expected
to show less contrast than CdSe since the former has fewer electrons per
unit cell. The line profile of the image contrast for Sample A and B
corresponding the the images are presented in Fig. 3 (b) and (d).  For 
Sample A, 
we observe a smooth drop in the
contrast near the edge of the nanocrystal. In comparison, 
the line profile of core-shell nanocrystals (Sample B), exhibit a trend of 
a stepwise-drop. The contrast can also be
explained by the change in thickness of nanocrystals, but it is not very
probable that it would occur in such a step-like way. From the HRTEM 
images the average shell thickness of the particles in Sample B has been 
estimated to 
be 1.2 nm.

From the micrograph [Fig. 3 a], we have
determined the lattice spacing in Sample A to be 3.72 \AA, which
corresponds to (100) plane of CdSe in hexagonal phase \cite{JCPDS}. 

\begin{table}
\begin{tabular}{|c|c|c|c|c|c|c|}\hline
{\bf Sample} & {\bf TEM} &  & {\bf OAS} & &{\bf Raman}\\ \hline
             & $\bar d$ nm & $\sigma$ & $\bar d$ & $\sigma$ & $\bar d$\\ 
\hline
Sample A & 5.2 & 0.22 & 6.0 & 0.22 & 4.7 \\ \hline
Sample B & 5.4 & 0.11 & 7.0 & 0.11 & 5.0 \\ \hline
\end{tabular}
\caption{Comaparison of the average size and size distribution of 
the particles in Sample A and B as obtained from blue shift in TEM 
images, optical absorption bands and confinement of phonons}
\end{table}

\section{Optical properties of nanocrystalline particles}

\subsection{Electron confinement :Optical absorption spectroscopy}

\begin{figure}
\centerline{\epsfxsize=4.5in\epsffile{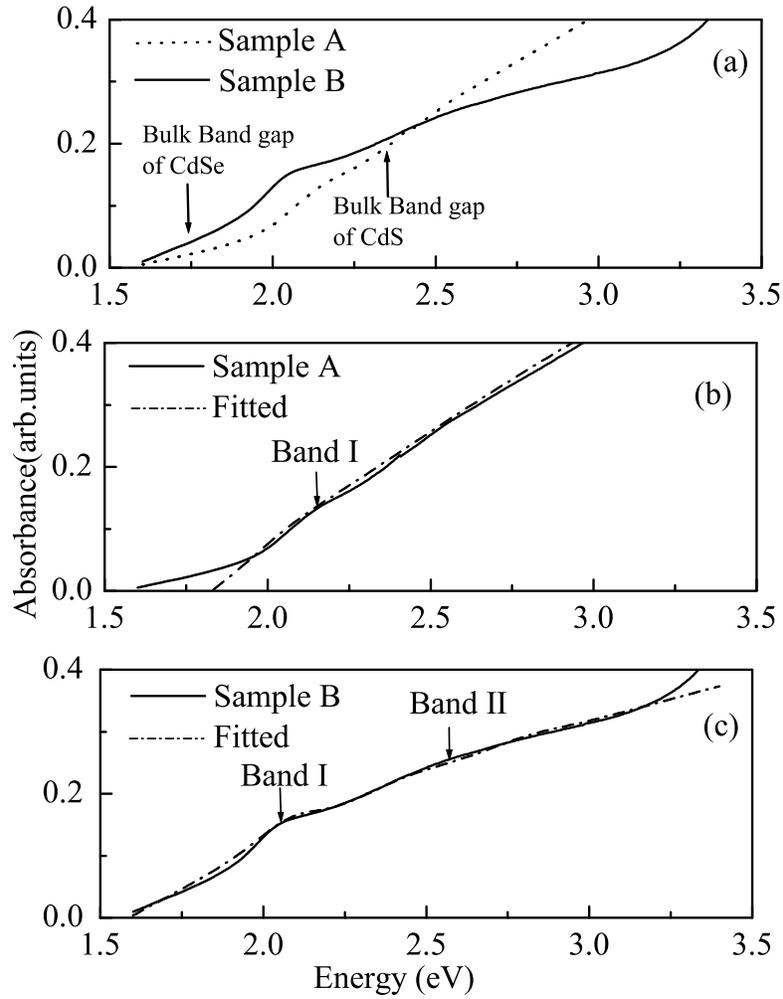}}
\caption{(a) Optical absorption spectra for Sample A and B. The dotted 
line is the spectrum for CdSe particles and solid line is for CdSe-CdS 
core-shell particles. The bulk band gap positions for CdS and CdSe are 
shown by arrows. (b) and (c) The dashed dotted lines are the 
nonlinear least square fit to the experimental curves (solid lines)  using 
effective bond order model.}
\end{figure}  

\begin{table}
\begin{tabular}{|c|c|c|c|}\hline
{\bf Sample} & {\bf Band I nm (eV)} & {\bf Band II nm (eV)} & {\bf PL peak 
nm (eV)}\\ \hline
Sample A & 577 (2.15) & -  & 538 (2.31)\\ \hline
Sample B & 606 (2.05) & 477(2.60) & 541 (2.29) \\ \hline
\end{tabular}
\caption{(a) Optical absorption and PL bands for Sample A and Sample B}
\end{table}

Fig. 4(a)  shows the optical-absorption spectra for the samples A and B. 
The absorption band energy (wavelength) from these samples are tabulated in 
Table II. The bulk band gaps for CdSe and CdS at room temperature are 1.74 
eV and 2.42 eV, respectively.
Samples A (CdSe particles) show only Band I at 2.15 eV in optical 
absorption spectrum (see Table II). We 
have observed that due to CdS shell growth a new band 
(Band II) at 2.60 eV appears in the optical absorption spectrum in 
addition to Band I. 

\begin{table}
\begin{tabular}{|c|c|c|c|c|c|c|}\hline
 & {Electronic transition} & {Confinement energy (ev)} & Type & f$_i$ & $ 
A_i$ & $x_i$  \\ \hline
CdSe & 1$\Gamma_{8}^{-}$-1$\Gamma_{6}^{+}$  & 0.42 & 1 & 4.85 & 55.50 & 
1.23 \\ \cline{2-6}
     & 1$\Gamma_{6}^{+}$-1$\Gamma_{7}^{-}$  & 0.83 & 2 & 1.11 & 38.71 & 
0.99 \\ \cline{2-6}
     & 1$\Gamma_{6}^{+}$-1$\Gamma_{8}^{-}$  & 0.84 & 2 & 0.10 & 38.71 & 
0.99 \\ \cline{2-6}
     & 2$\Gamma_{7}^{-}$-1$\Gamma_{6}^{+}$  & 0.72 & 3 & 0.61 & 70.35 & 
1.17 \\ \cline{2-6}
     & 2$\Gamma_{8}^{+}$-1$\Gamma_{8}^{-}$  & 0.75 & 3 & 1.89 & 70.35 & 
1.17 \\ \cline{2-6}
     & 1$\Gamma_{7}^{+}$-1$\Gamma_{8}^{-}$  & 0.78 & 3 & 0.62 & 70.35 & 
1.17 \\ \cline{2-6}
     & 3$\Gamma_{7}^{-}$-1$\Gamma_{6}^{+}$  & 0.76 & 3 & 0.01 & 70.35 & 
1.17 \\ \hline
CdS  & 1$\Gamma_{8}^{-}$-1$\Gamma_{6}^{+}$   & 0.40 & 4 & 4.13 & 45.40 & 
1.19 \\ \cline{2-6}
     & 1$\Gamma_{7}^{-}$-1$\Gamma_{6}^{+}$   & 0.45 & 4 & 1.73 & 45.40 & 
1.19 \\ \hline
\end{tabular}
\caption{The confinement energy, $\Delta E$, changes with diameter of the 
particle $d$ by following the relation $\Delta E = A_{i}/d^{x_i}$; where 
$\Delta E$ is in eV and $d$ in \AA}
\end{table}  

The blueshift in the optical absorption spectra of the particle
from that for bulk CdSe and CdS arises from the confinement of charge
carriers in the nanocrystals.  Assuming the particles to be spherical, the
optical-absorption coefficient of the collection of the monodispersed
particles of average diameter $\bar{d}$ at low temperature is given by

\begin{equation}
\alpha(E)=\sum\frac{f_{i}\Gamma_{i}}{(E-E_{i})^{2}+\Gamma_{i}^{2}},
\end{equation}

\noindent
where, $f_{i}$ is the oscillator strength, $E_{i}$ is the transition 
frequency and $\Gamma_{i}$ is the half width at  half maxima (HWHM) for 
the 
$i$th interband transition. Due to  thermal broadening and inhomogeneous 
broadening due to size distribution of the particles Eqn. 2 is modified 
to 

\begin{equation}
\alpha_{observed}(E)=B\sum_{i}\int_{0}^{\infty}d(d)\frac{P(d)f_{i}\Gamma_{i}}
{[E-E_{i}(d)]^{2}+\Gamma_{i}^{2}},
\end{equation}

\noindent
where $P(d)$ is the log-normal size distribution of the particles, same as 
Eqn.1. The results of single band effective mass approximation model do 
not 
agree with the optical absorption spectra of the nanocrystals because it 
neglects intervalence band mixing and deviation from quadratic 
dispersion. 
This has been shown in the literature for bare CdSe or CdS particles and 
CdS-CdSe mixed crystals \cite{Ramaniah:1993,Roy:1996}. We have 
assumed results from 
effective bond order model by Ramaniah and Nair on interband 
transition in quantum dots of CdS and CdSe and considered the 
following transitions : 
1$\Gamma_{8}^{-}$-1$\Gamma_{6}^{+}$, 1$\Gamma_{6}^{+}$-1$\Gamma_{7}^{-}$, 
1$\Gamma_{6}^{+}-$1$\Gamma_{8}^{-}$, 2$\Gamma_{7}^{-}$-1$\Gamma_{6}^{+}$, 
2$\Gamma_{8}^{+}$-1$\Gamma_{8}^{-}$, 1$\Gamma_{7}^{+}$-1$\Gamma_{8}^{-}$, 
3$\Gamma_{7}^{-}$-1$\Gamma_{6}^{+}$, 
for CdSe core 1$\Gamma_{8}^{-}$-1$\Gamma_{6}^{+}$, 
$\Gamma_{7}^{-}$-1$\Gamma_{6}^{+}$ for CdS shell \cite{Ramaniah:1993}.  
Confinement energies for these transitions as obtained in Ref. 
\cite{Ramaniah:1993}  are tabulated in Table III. 
We have grouped these transitions in 3 types for CdSe and 1 type for CdS, 
according to their confinement energies. Since the oscillator strength is 
not very sensitive to particle diameter $\bar{d} \geq 2-3$ nm, 
the value of $f_{i}$ corresponds to those of $d$= 4.8 nm for 
transitions in  CdS and $d$= 4.6 nm  for CdSe as obtained in Ref. 
\cite{Ramaniah:1993}.  The 
confinement energy ($\Delta$E) for these 4 types of transitions $vs.$ 
particle size have been plotted in Fig. 5.  For $\Delta(E)$
we have assumed the empirical relation \cite{Roy:1996}

\begin{equation}
\Delta(E)=A_{i}/d^{x_{i}},
\end{equation}

\begin{figure}
\centerline{\epsfxsize=4.5in\epsffile{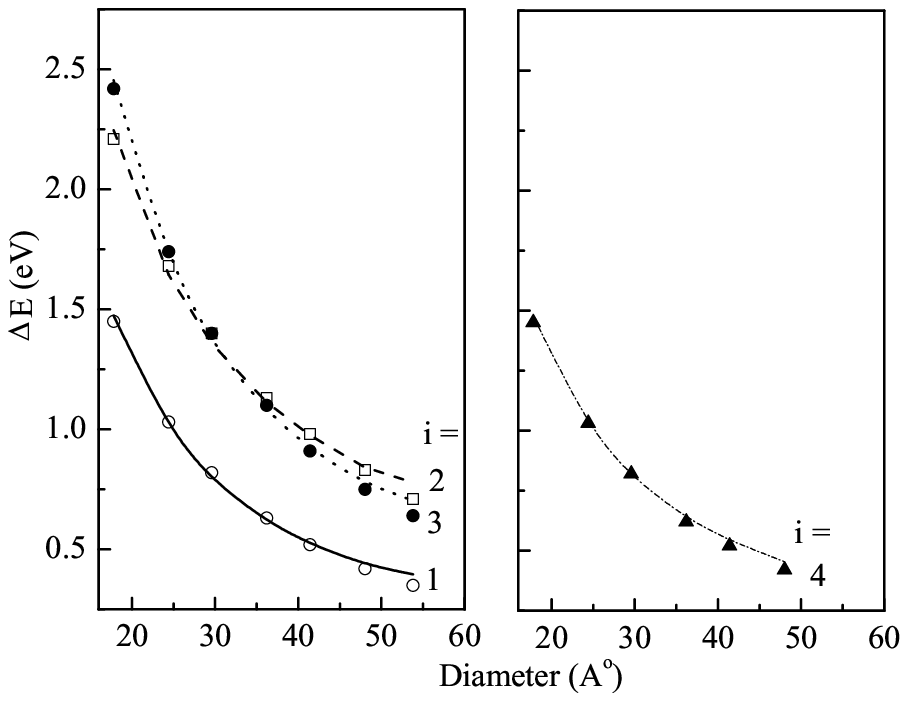}}
\caption{The variation in $\Delta$E with respect to diameter, $d$, of the 
particle for different transitions in effective bond order model for CdSe 
and 
CdS. The curve through the points are fit to $\Delta E=A_{i}/d^{x_{i}}$.}
\end{figure}  

\noindent
where $A_{i}$ and $x_{i}$ are constants for a particular transition. 
We have  fitted the 4 types of transitions with Eqn.4 (shown in Fig. 5) 
and estimated the values of $A_{i}$ and 
$x_{i}$ for these transitions, which are given in Table III. Taking 
$E_{i}(d)=E_{g}^{bulk}+A_{i}/d^{x_{i}}$ we have fitted the 
experimental optical absorption
spectra, shown by dashed-dotted lines in Fig 4(b) and (c) for Sample A 
and B, respectively. It can be seen that  the
theoretical curves fit the experimental data reasonably well. The
values of average particle size and $\sigma$ from the above analysis are 
tabulated in Table I. The same Table also reveals the 
comparison between the average sizes of the particles  obtained from 
optical absorption measurements and HRTEM  frequency plots.

\begin{figure}
\centerline{\epsfxsize=4.5in\epsffile{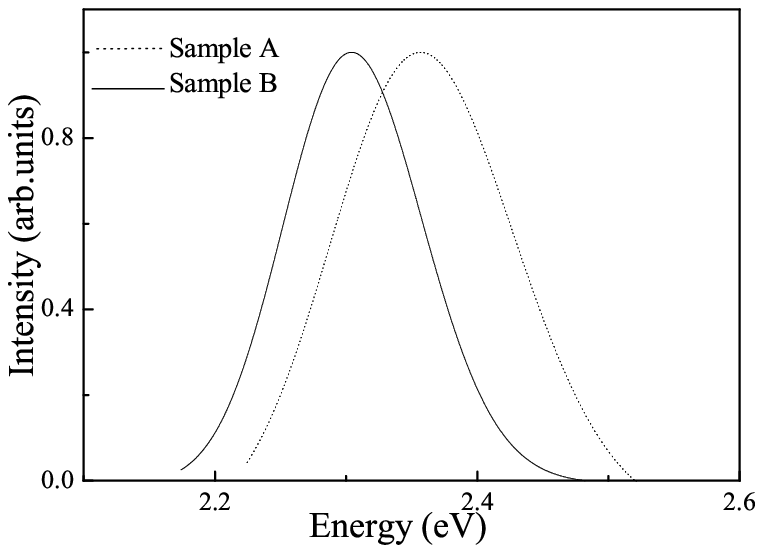}}
\caption{PL spectra for Sample A and B.}
\end{figure}

\subsection{Photoluminescence}

Photoluminescence spectra is obtained using 1200 g/mm holographic 
grating, 
a holographic supernotch filter, and a Peltier cooled CCD detector. 
Spectra 
are taken using 488 nm Argon ion laser as an excitation source . 

Figure 6 shows the PL spectra from the Samples A and B. The PL peak 
positions, obtained by fitting each spectrum by a Gaussian distribution 
have  been tabulated in Table II.
PL peaks from all the samples are slightly 
blue or red shifted from the first excitonic peak in the absorption 
spectra.  This shift results from the convolution of the size 
distribution and the emitting state and excitation energy for the 
excitation near the first peak \cite{Hoheisel:1994,Nirmal:1995}. 
The detail study on the photo-stability of the particles, prepared exactly 
by 
the same route, has been reported in Ref \cite{Liu:2000}.

\section{Confinement of optic phonon and new surface phonons\\:Raman 
scattering}

\begin{figure}
\centerline{\epsfxsize=4.5in\epsffile{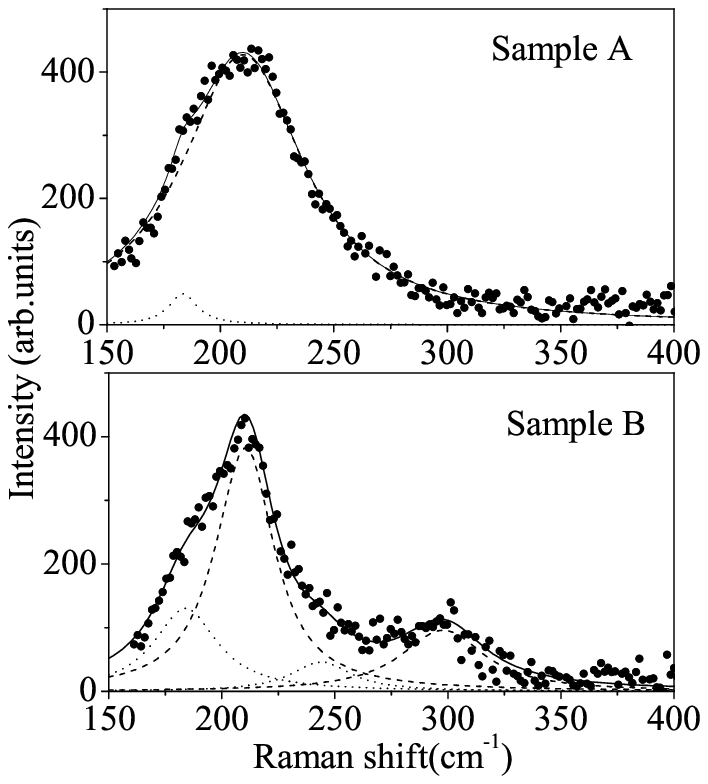}}
\caption{Raman spectra for Sample A and Sample B are shown by filled 
circles. Phonon confinement 
components and surface phonon components are given by dashed and dotted 
lines, respectively. Solid lines correspond to the best fit to the 
experimental data using combined Raman line shapes} 
\end{figure}  

In order to understand the nature of phonon confinement and appearance of 
new surface modes in our samples, we have performed Raman scattering 
studies.
Raman spectra are obtained using the same experimental set up as used for 
the above mentioned PL measurements. 
The slit width of the spectrometer during the experiment was 50 $\mu$m.
The first order Raman spectra for the 
samples are shown in Fig. 7. Using the phonon confinement model of
Campbell and Fauchet \cite{Campbell:1986}, the first order Raman spectrum 
$I_{c}(\omega)$ is given by

\begin{equation}
I_{c}^{j}(\omega)=A \int_{0}^{q_{max}}\frac{d{\bf q}{\vert C(0,{\bf 
q})\vert}^{2}}{\left[\omega-
\omega({\bf q})\right]^{2} + \left(\Gamma_{0j}/2\right)^{2}},
\end{equation}                     

\noindent
where  $\omega({\bf q})$ and $\Gamma_{0j}$   are the
phonon dispersion curve and the natural line width (FWHM) of the
corresponding bulk materials, $C(0,{\bf q})$   is the Fourier coefficient
of  the phonon confinement function. $A$ is an arbitrary constant. $j$=1 
for CdS and $j$=2 for CdSe. For
nanoparticles, it has been shown that the phonon
confinement function, which fits the experimental data best, is
$W(r, \bar d_{z})=\exp(\frac{-8\pi^{2}r^{2}}{\bar d^{2}_{z}})$,
the square of the Fourier coefficient of which is given by ${\vert
C(0,{\bf q})\vert}^{2} \cong \exp\left({-\frac{q_{2}^{2} \bar 
d^{2}_{z}}{16\pi^2}}\right)$. 
Here, $\bar d_{1}$ is the average size of the spherical nanocrystals in 
Sample 
A and $\bar d_{2}$ is the average diameter of the core in Sample B. On the 
other hand, for the shell
component of particles in core-shell structure (Sample B), we have used an
additional phonon confinement function for CdS shell, given by
\cite{Campbell:1986}

\begin{equation}
{\vert C(0,{\bf q})\vert}^{2} \cong \exp\left(-\frac{q_{1}^{2}
t^{2}}{16\pi^2}\right)
{\left\vert
1-erf\left(\frac{q_{1}t}{\sqrt{32}\pi}\right)\right\vert}^{2}
\end{equation}

\noindent
where $t$ is the average shell thickness for the
particles. The average 
phonon dispersion in the bulk CdSe and CdS crystal for the LO phonon 
modes are taken as \cite{Cardona:1998,Bliz:1979}

\begin{equation}
\omega(q)=\omega_{0}^{j}-\Delta\omega^{j} (q_{j}^{2})
\end{equation}

\noindent
which fits the experimental curve well in the direction of
$\Gamma$-M upto $q_{max}$ = 0.4.
$\omega_{0}^{j}$ is the corresponding bulk LO phonon frequency 
:$\omega_{0}^{^1}$ = 302 cm$^{-1}$ and $\omega_{0}^{^2}$= 213 cm$^{-1}$.
$\Delta\omega^{j}$ is the band width of bulk LO phonon branch :
$\Delta\omega^{1}$ = 102 cm$^{-1}$, $\Delta\omega^{2}$ = 118 cm$^{-1}$. 
In the above equations, $q_{j}$ is taken in unit of $2\pi/a_{j}$, where,
$a_{j}$ is the lattice constant of the material. We 
have taken $a_{1}$=5.82 \AA \ and $a_{2}$=6.08 \AA \ \cite{JCPDS}. 

In addition to LO phonon peak, we have also observed additional peaks in 
the spectra for Sample A and  B. 
Keeping in mind the possibility of the presence of the surface phonons 
(SP) between longitudinal and transverse optical phonon modes, together
with the confined optical phonon, we have fitted this 
additional peak with the Lorentzian function
\begin{equation}
I_{SP}^{j} (\omega)= 
\frac{B\Gamma_{SP}}{(\omega-\omega_{SP})^{2}+\Gamma_{SP}^{2}},
\end{equation}

\noindent
where $\omega_{SP}$ and $\Gamma_{SP}$ are the peak position and  the HWHM, 
respectively, for the SP mode. We fit the full spectrum for Sample A and B 
by a combined line shape $I(\omega) = 
I^{j}{_c}(\omega)+I^{j}_{SP}(\omega)$. The best fit obtained for both the 
samples are shown by
the solid lines in Fig. 7. The phonon confinement components are shown by 
dashed lines 
and the surface phonon components are shown by dotted lines
in Fig. 7. In the fitting procedure, we have kept $\bar d_{z}$, $t$,
$\omega_{SP}$, $\Gamma_{SP}$, $A$ and $B$ as fitting parameters. For the 
best fit of the spectra the
values of $\Gamma_{02}/2$ for Sample
A and Sample B are taken to be 32 cm$^{-1}$ and 14.5 cm$^{-1}$.
We attribute the variation in the value of
$\Gamma_{02}/2$ from
sample to sample to the difference in size distribution of the
particles. From the non-linear least square fit, the average particle 
size ($\bar d_{1}$) for Sample A is obtained as 4.7 nm.
The core diameter ($\bar d_{2}$)and 
shell (film) thickness, $t$,  for Sample B are obtained as 4.0 nm and  
1.25 nm. This estimates average particle diameter ($\bar d_{2}+t$) in 
Sample B to be 5.25 nm. The 
diameters of the particles in Sample A and B thus measured are 
very close to the same estimated from optical absorption and
HRTEM measurements [see Table I]. We have observed  
that the SP modes for Sample A at 183 cm$^{-1}$ and for CdSe and CdS -like 
modes in  Sample B at 180 cm$^{-1}$ and 244 cm$^{-1}$, respectively. 
The ratio of $A/B$ is 2$\times$10$^3$ for Sample A, 
whereas, the ratio of the same for CdS-like and CdSe-like modes in sample B 
are 85 and 68.

For core-shell nanocrystals, we have considered the optical
modes as the response of the two types of oscillators under applied
electric fields. The effective oscillator strength is proportional to the
density of the oscillators. Assuming nonoverlapping restrahlen
bands for these two modes  
$\omega_{LO1}>\omega_{TO1}>\omega_{LO2}>\omega_{TO2}$, the
dielectric response function of a core-shell particle can be taken as 
\cite{Roy:1996}

\begin{equation}
\epsilon(\omega)=p\epsilon_{1}(\omega)+(1-p)\epsilon_{2}(\omega)
\end{equation}

\noindent
where $\epsilon_{1} (\omega)$ and  $\epsilon_{2} (\omega)$ are given by

\begin{equation}
\epsilon^{j}(\omega)=\epsilon_{\infty}^{j}\left[1+\frac{\omega_{LOj}^{2}-      
\omega_{TOj}^{2}}{\omega_{TOj}^{2}-\omega^2}\right]
\end{equation}

\noindent
and $\epsilon_{\infty}$ is the high frequency dielectric constant of the
crystal. Here, we have taken $p=t/(t+x)$, where $t$ is the CdS shell 
thickness, as taken earlier and $x$ is the depth of penetration of the 
surface phonon within the CdSe core region of the core-shell structure. 
Substituting Eqn. 10 in Eqn. 9 for both CdS and 
CdSe like modes and
using $\epsilon(\omega)=-(l+1)\epsilon_{m}/l$ \cite{Fuchs:1968}, we get 
the following equation for the surface phonon frequencies of the 
core-shell nanocrystals:   

\begin{eqnarray}
\omega^{4}-\omega^{2}\left[\omega^{2}_{TO1}+\omega^{2}_{TO2}+\frac
{p\epsilon_{\infty}^{1}}{K}(\omega^{2}_{LO1}-\omega^{2}_{TO1})
+\frac{(1-p)\epsilon^{2}_{\infty}}{K}(\omega^{2}_{LO2}
-\omega^{2}_{TO2})\right]\nonumber \\
+\omega^{2}_{TO1}\omega^{2}_{TO2}\left[1+\frac{p\epsilon_{\infty}^{1}}{K}
\frac{(\omega^{2}_{LO1}-\omega^{2}_{TO1})}{\omega^{2}_{TO1}}
+\frac{(1-p)\epsilon_{\infty}^{2}}{K}\frac{(\omega^{2}_{LO2}-\omega^{2}_{TO2})}
{\omega^{2}_{TO2}}\right]=0
\end{eqnarray}    

\noindent
Here,
$K=(l+1)\epsilon_{m}/l+p\epsilon^{1}_{\infty}+(1-p)\epsilon_{\infty}^{2}$;
$\omega_{LOj}$ and $\omega_{TOj}$ are longitudinal and transverse mode
phonon frequencies for CdS and CdSe. $\epsilon_m$ is the dielectric 
constant of the medium, marcaptoacetic acid in our case. The theory of 
electron-phonon 
interaction shows that SP  modes with $l\geq$ 1 and those with $l$=even 
can contribute to Raman scattering via the deformation potential of 
electron-phonon interaction \cite{Ruppin:1970,Fedorov:1997} or the 
Fr\"{o}hlich electron-phonon 
coupling\cite{Comas:2002}, respectively. The higher SP modes ($l>2$) 
contribute 
only slightly to the scattering intensity. Thus, we can assign the 
surface modes only with $l$=2. As before, we have taken the values 
of $\omega_{L01}$ and $\omega_{LO2}$ as 302 cm$^{-1}$ and 213 cm$^{-1}$ and the 
values of $\omega_{TO1}$ and $\omega_{TO2}$ as 238 cm$^{-1}$ and 168 
cm$^{-1}$, respectively. The values of other parameters are
$\epsilon_{m}$ =14.3,
$\epsilon_{\infty}^{1}$=5.5 and $\epsilon_{\infty}^{2}={8.9}$ \cite{CRC}.
The average value of $t$ is known from HRTEM and Raman measurements to 
be 1.2 nm.  The solid line and the dashed line in Fig. 8 
shows the variation of CdS-like and CdSe-like SP frequencies with $x$ for 
$l=2$ using Eqn. 11. 
The observed values of CdS-like and CdSe
like SP frequencies in Sample B are shown as filled circles in Fig. 8. 
These values correspond to $x$= 1.8. 

\begin{figure}
\centerline{\epsfxsize=4.5in\epsffile{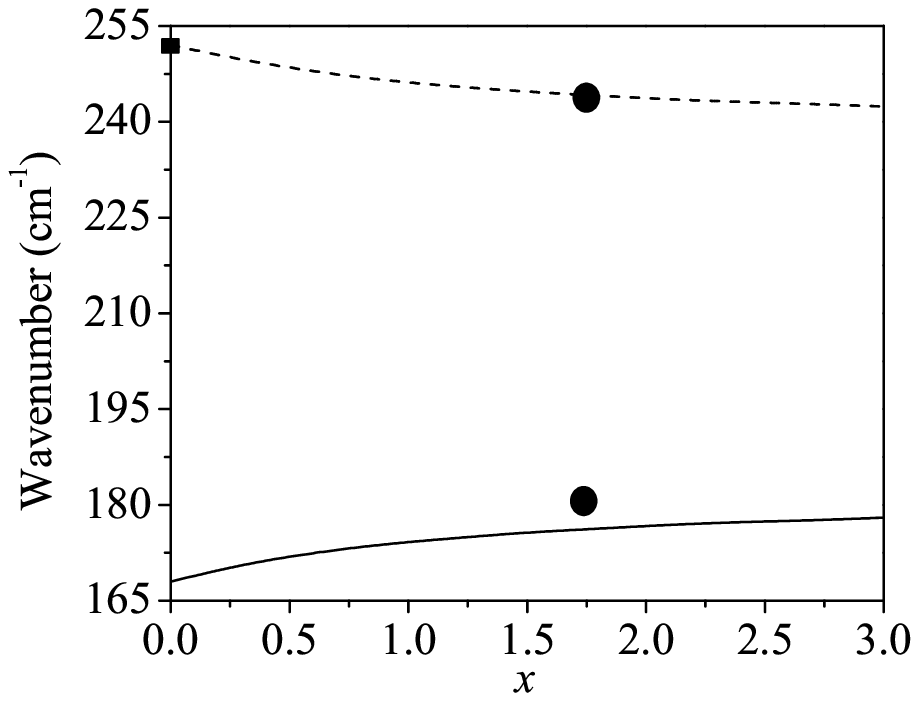}}
\caption{Variation of CdS like (dashed line) and CdSe like (solid line) SP 
phonon frequencies with $x$. The solid circles are SP frequencies for the 
Sample B. The solid square is the estimated SP frequency in 
pure CdS nanoparticle [$x$=0 ; $p$=1]}
\end{figure}  

We have observed that the SP mode for 
Sample A at 183 cm$^{-1}$, which is very
close to the value of $\omega_{SP}$ (182 cm$^{-1}$) for pure CdSe 
particles, obtained from Eqn. 10. On the other hand, the observed CdS-like 
SP frequency  (244 cm$^{-1}$) from the shell component of the Sample B is 
away from the frequency (252 cm$^{-1}$) of the same for 
pure CdS particles, estimated directly from Eqn. 10 or with $x$=0 ($p$=1) 
in Eqn. 11 (shown by filled square in Fig. 8). 
However, it is  close to the value, 243 cm$^{-1}$, obtained from 
Eqn. 12 with $x$=1.8 using dielectric response function model for two 
oscillators. The corresponding CdSe-like mode in Sample B is expected to 
appear at 176 cm$^{-1}$. The mismatch between this estimated value and 
observed frequency of SP mode (180 cm$^{-1}$) for the core material may be 
due to the effect of interface between core and shell structure. 
The SP frequencies for the CdS-like
and CdSe-like modes in Sample A and B as obtained from the experiments and
above calculations are tabulated in Table IV.
                                              
It is interesting to note that if CdSe-CdS would form a mixed crystal in 
the solution, 
we expect (from detailed theory of SP modes in mixed crystals) CdSe-like
and CdS-like SP modes for sample B to appear at 204 cm$^{-1}$ and  267 
cm$^{-1}$, respectively \cite{Roy:1996}. This is far away from 
what we have got experimentally. Thus, the above discussion implies that 
from the 
SP frequencies of the expected shell material, one can  confirm the  
formation of core-shell like particle, rather than formation of mixed 
nanocrystallites or individual particle of shell material in 
the sample. 

\begin{table}
\begin{tabular}{|c|c|c|c|c|}\hline
{\bf Sample} & {\bf $\omega_{SP}$CdSe(calc)}& {\bf 
$\omega_{SP}$CdSe(expt)}
& {\bf $\omega_{SP}$CdS(calc)} &  {\bf $\omega_{SP}$CdS(expt)} \\ \hline
Sample A & 182  & 183 & - & - \\ \hline
Sample B & 177 & 180 & 243 & 244 \\ \hline
\end{tabular}
\caption{Calculated and experimental CdS and CdSe- like SP frequencies in 
Sample A and B. The unit for above frequencies are in cm$^{-1}$}
\end{table}

\section{Conclusion}
Our goal in this article was to show electron and phonon confinement in 
the CdSe-CdS core-shell nanostructure. To this end, we have first 
confirmed the existence of the core-shell structure in our samples, 
prepared by soft chemical route, using HRTEM images. Subsequently, to 
analyze the  electronic transitions we focus on optical absorption 
measurements, supported by theoretical considerations of the effective 
bond order model. Furthermore, the first order Raman line shape of the 
CdSe-CdS nanocrystals is quantitatively explained by taking into account 
both confined phonon modes and  SP modes. The frequencies of SP modes 
are shown to match well with their calculated values as obtained from the 
dielectric response function 
model. We also demonstrate that due to passivated surface states CdS 
capped samples exhibit clear SP modes separated from LO modes in core-shell 
type nanocrystallites. In summery, we have presented a technique to 
confirm the core-shell structure in this type of semiconductor nanostructure.

While analysing the optical absorption of the CdSe-CdS
core-shell structure using effective bond order model, we have assumed the 
expected electronic transitions for CdSe and CdS particles. The phonon 
confinement in the shell structure is explained by 
taking the shell as a thin layer on CdSe core. The above assumptions may 
or may not play a crucial role in analyzing the data. However, more 
accurate theoretical  models are necessary to understand core-shell 
strucure better.

\section{Acknowledgement}
AR thanks DST and BRNS in India, for financial support.


\end{document}